\documentclass[preprintnumbers,amsmath,amssymb]{revtex4}
\usepackage{graphicx}% Include figure files
\usepackage{dcolumn}% Align table columns on decimal point
\usepackage{bm}% bold math

\begin{document}

\preprint{YITP-05-38}
\preprint{OIQP-05-08}
%\date{October 1, 2005}

\title{Intrinsic Periodicity of Time and Non-maximal Entropy of Universe}
% Force line breaks with \\

\author{Holger B. NIELSEN}
\affiliation{%
Niels Bohr Institute\\
17, Blegdamsvej, Copenhagen\\
Denmark
}%

\author{Masao NINOMIYA}
 \altaffiliation[Also at ]
 {Okayama Institute for Quantum Physics, Kyoyama-cho 1-9, 
 Okayama City 700-0015, Japan.}%Lines break automatically or can be forced with \\
%\author{}%
\affiliation{
Yukawa Institute for Theoretical Physics,\\
Kyoto University\\
Kyoto 606-8502, Japan% with \\
}%

\begin{abstract}
The universe is certainly not yet in total thermodynamical equilibrium,
so clearly some law telling about special initial conditions is needed. 
A universe or a system imposed to behave periodically gets thereby
required ``initial conditions".
Those initial conditions will \underline{not} look 
like having already suffered the heat death, i.e. obtained the maximal
entropy, like a random state.
The intrinsic periodicity explains successfully why entropy is not
maximal, but fails phenomenologically by leading to a \underline{constant} entropy.
% \verb+\pacs{#1}+ command.
\end{abstract}

%\pacs{11.10, 98.80.E, 98.80.C, 05.07}% PACS, the Physics and Astronomy
\maketitle

\section*{Introduction}
In general relativity, one is forced to accept the existence of singularities 
which represent points at which time stops or starts under quite possible
conditions \cite{1}. 
However, today a priori it is not excluded that the true laws of nature
could be so as to have no such singularities.
In such a case one would have to choose among options such as:
\begin{enumerate}
  \item Universe has always existed and will always exist.
  \item Universe returns periodically to the exactly same state after some period $T$.
  \item Time is not simply a real variable as we are accustomed to, but e.g. 
the imaginary time is effectively considered in Hartle-Hawking' 
no boundary quantum cosmology
\cite{2}.
\end{enumerate}

It is the purpose of the present article to discuss the second possibility
of an intrinsically periodic universe and to connect that with the questions
of the ``initial state" of the universe and the second law of thermodynamics.

It is well-known that the second law of thermodynamics requires that back in
time the world were in a very special state.
Then as it develops more and more of the original order on the micro level
gets spoiled and we may conclude that entropy increases \cite{3}\cite{4}\cite{5}.

For the purpose of the present article we want to separate out
a couple of mysteries caused by the second law of thermodynamics as seen
in the light of the other laws from the microphysics point of view.
With a point of view that everything unsettled by the equations
of motion should be taken a priori to be random we can claim some mysteries:
\begin{enumerate}
  \item Why should the universe not be in the state of total thermodynamical 
equilibrium -- the heat death -- ? After all, by far the largest phase space volume of 
a mechanical system corresponds effectively to the system in the maximal entropy state, 
the heat death macro state.
  \item It is even more mysterious that the state in early times of universe 
were much more special --i.e. it had much lower entropy -- than today.  
A law that should be imposed to such special state of the dynamical variables 
seems strangely enough time translational invariant. 
In fact it does not to impose the same entropy at different times.
  \item In the light of the other laws which have 
with rather high accuracy time reversal invariance, 
it is even more mysterious that the great law about the entropy 
-- the second law of thermodynamics -- is totally in contradiction 
with this time reversal symmetry. 
This is the arrow of time problem
\cite{7}\cite{8}.
\end{enumerate}

It is one of the points in the present article \cite{6} 
to put forward and stress that the following point 2 above 
-- letting the universe return with a given period to exactly the same state, 
fixes the initial conditions.
So with an $S^1$-compactified time axis -- which is periodic -- 
there is ``essentially" no more place for imposing initial conditions.
That is to say that with the $S^1$-compactified time axis 
we have no more any freedom to put in restrictions on the initial 
conditions so as to make a theory behind the second law of thermodynamics.
In such an $S^1$-time world one rather gets the second law 
or one does not get it as calculations, below, may show.

It is actually the point of the present article that with an $S^1$-time 
-- i.e. intrinsic periodicity -- 
one does \underline{not} get the second law of thermodynamics 
in a trivial way $\dot{S}=0$! 
But and this is at least something:  
The $S^1$-time model solves the first one of the above listed mysteries about entropy:  
It explains that it is very likely that entropy is not maximal !

%\subsection{\label{sec:level2}Second-level heading: Formatting}

\section*{Macro State Consideration}

It is well-known how describes a given macro state by a Gibbs ensemble one 
in statistical mechanics, which really means a probability distribution 
over phase space for the system conceived of degrees in terms of micro degrees of freedom 
(the fundamental physics description).
For the practically most important systems the number of degrees of freedom 
is so huge that within the needed accuracy we in fact can obtain the macro variables 
characterizing the macro states from just \underline{one} typical micro degree of freedom, 
i.e. a point in the phase space. 
For instance just one single -- in classically thinking -- 
state of the molecules in a set of containers containing gases of different compositions, 
temperatures and pressures could tell us the energies, 
the number of molecules of the different sorts in the different containers.
Now for the systems that are big enough 
we can approximate the micro canonical ensembles with canonical ones and vice versa, 
and grand canonical ensembles with
systems with given number of particles and vice versa.
Thus from just knowing the numbers of molecules of different sort and the total energy of the molecules 
in each of the containers we can, estimate with sufficient accuracy, 
the chemical potentials and temperatures in all the containers.
So from just knowing one micro state we can derive the macro state. 
If it were not so we could not measure the temperature in a given situation 
but could only define say the temperature for an ensemble of cases.

Since there corresponds a macro state to micro states 
(= phase space points (q, p)) we can consider the macro states 
subsets of the phase space.

In that point of view the history or development of the universe 
is fundamentally given by the development of a micro state which is just a phase space point 
(we ignore totally quantum mechanics in the first part of this article).  
Then as time goes on it passes through various macro states, 
which are subsets it meets on its way.

It is well known that looked at this way the entropy $S(A)$ of a macro state is 
essentially the logarithm of the phase space volume.
Here we say ``essentially" it is because of the following reasons: \cite{9}
\begin{enumerate}
  \item If as it is typically the case the macro states 
are characterized by real variables such as energies 
in various containers or momentum or angular momentum in them or positions of pistons, 
then the phase space subset corresponding to such macro states have, 
strictly speaking, lower dimension than the phase space itself.
Thus they do have zero phase space volume. However, 
within the accuracy needed a ``reasonable" smearing out to a non-zero volume subset 
could be done effectively un-ambiguously, 
although in principle the entropy would have a logarithmic dependence, 
but in practice insignificant dependence on the smearing out scale.
  \item Usually you may define the entropy from the statistical distribution of a Gibbs ensemble 
rather than from the point of view of macro states as subsets used here. 
But again this is not significant within the accuracy achieved in macro experiments.

We may well assume that one can define macro states for example by how energy momentum etc. 
is distributed into different ``containers" or parts of the system, 
and thus also determine an entropy from the just mentioned logarithm 
of the phase space volume of the subset associated.
However in the present article we shall not assume the second law of thermodynamics as valid.  
Rather the philosophy of this article is to speculate over 
if we could make a model behind this second law o f thermodynamics so that we could derive it.
Unfortunately that turns out to be unsuccessful and we only get rather weak result of the entropy;  
not likely maximal as one might a priori fear easily could happen.
In such an approach we shall not only at first ignore the second law of thermodynamics, 
but also throw away those usual intuitions and assumptions associated with it as well as the time arrow
\cite{7}\cite{8}.
Once we throw away this second law and the associated prejudices 
there is well known from micro physics full time reversal invariance, 
except for the tiny effects associated with Yukawa coupling matrices known from CP-breaking first in 
$K^0$-decay.
Thus without these assumptions there is nothing strange 
in the milk and the black coffee separating out of the brown mixture.
A priori we can have the micro state -- walk the phase space point $(q, p)$ --
as function of time from one macro state to another, e.g. 
through the macro state with the mixture of milk and black coffee to the macro state 
where they are separated and may be back again to the mixture or some totally different macro state.

If we just use the micro physical laws -- in our classical treatment --, 
simply the Hamilton-equations, many such developments back and forth between macro states 
are logically and equation of motion-wise possible
what really can happen depends on the ``initial conditions". 
These ``initial conditions" are just meant to be specification of the solution by giving e.g. 
the phase space point at some moment of time $t_0$ which does not have to be in any sense ``initial".  
\end{enumerate}

\section*{Searching for solutions}
In general one could imagine searching for solutions to the equations of motion 
-- for instance for solutions with a given period $T$ as is the main study of the present article -- 
by considering a start at some time $t_0$ say and then integrate up from that time perhaps 
even both forward and backward in time and then investigate the properties, 
i.e. most importantly through which macro states it will go. 
In principle you can imagine performing such a calculation for all points in phase space 
taken to be the one passed at time $t=t_0$, and you can ask for the phase space volume or rather 
its logarithm for the set of all the phase space points $(q(t_0), p(t_0))$ 
leading to a solution passing through a prescribed series of macro states at various moments of time.
In principle one may fear that the phase space volume for micro states 
realizing such a macroscopically prescribed development will depend 
much on the actual Hamiltonian of  the system of macro states. 
However if we make a few rather weak and reasonable assumptions 
we can with sufficient accuracy estimate the logarithm 
of the phase space volume of the set of the solutions with a given macro history.

The mild assumptions we shall make are the following:
\begin{enumerate}
  \item Whenever we postulate that the solutions counted 
by their phase space volume is $t_0$ pass at a moment of time 
from one macro state to the next one, we only discuss series of macro states 
so that the two macro states involved are what we can call ``maximally coupled".

By this ``maximally coupled" we mean that provided a ``negligible number" of degrees of freedom 
are put properly all phase space for one of the macro state develops 
in to the other one either backward or forward in time.
By this concept of a ``negligible number" of degrees of freedom 
we would typically have in mind a set of macro variables such as the stand of a value say, 
making an opening between two containers.
If the value is indeed open -- 
requiring only one degree of freedom to have the appropriate value, 
and one is indeed negligible compared to say Avogadro's number 
-- all the micro states with the molecules in one of the containers can indeed -- 
in fact both backward and forward -- go into the state with the molecules in both containers.
  \item  The second assumption is that we consider a generic system.
We could make this statement precise by thinking that we take the Hamilton part, 
which conserve the negligibly few macro degrees of freedom to be essentially random.
That should mean that how the phase space point $(q(t)_i ~p(t)_i)$ 
runs around inside the various macro states which comes through 
-- conceived of as subsets -- is largely given by this random part of the Hamiltonian 
and can effectively be considered in a purely statistical way.

Really, under the just proposed assumptions for a time-series of ``maximally coupled" 
macro states we want to derive the following rules 
for the logarithm of the phase space volume of the solutions going through this series:
\begin{enumerate}
  \item Only requiring the system to pass one macro state 
can be achieved with the logarithm of the phase space volume equal to this logarithm 
of volume for just that macro state. 
But that is just the entropy of that macro state. 
So
\begin{eqnarray}
\log {\rm Vol}(``macro ~state~  A")=S(A)
\end{eqnarray}
where $S(A)$ is the entropy of the macro state A.

We add to our assumptions that we only consider what we could call 
``reasonable times" meaning that they are small in relevant units compared to exponentials of entropies. 
  \item When the series of macro states contain neighboring macro states in the time series with different entropy 
it is unavoidable that at most the fraction
$e^{S_{small}-S_{big}}$ of the solutions passing through the volume wise bigger macro state -- 
with entropy called $S_{big}$ -- can also come through the smaller one -- 
having entropy $S_{small}$ --.
This kind of statistical consideration that we can treat such estimates 
as giving a chance for each solution is justified by the random Hamiltonian philosophy assumed here.
  \item Our rule for a (long) chain of macro states for the phase space volume 
that can get through this long chain of macro states, then becomes the following:

Ignore first those outer-most -- meaning first in time or last -- 
which have bigger entropy than the successively more inner one, 
and continue to do so until we are left with the now outer-most macro states having 
less entropy than their respective successor and predecessor for the first and the last.

Then for this reduced chain the phase space volume is the same as 
for the original chain of macro states. 
In fact it is given by
\begin{eqnarray}
\exp\left(
\sum _{
\substack{
i \in minimal ~S\\~macro ~states}}
S_{min.i}
-
\sum _{\substack{j \in maximal ~S\\~macro ~states}}S_{max.i}
\right)
\label{2}
~\\
\mbox{with } i \in \left\{minimal-S ~macro ~states \right\}
~. \nonumber 
\end{eqnarray}
Here we have looked along the chain and selected the macro states that 
have minimal entropy $S_{min.i}$ in the sense of having lower entropy 
than both foregoing and successive macro states in the chain.
The first term in the exponent is the sum over the entropies of these macro states.
Next one analogously selects those macro states in the reduced chain 
which have bigger entropy than both its neighboring macro states, 
and the second term is the sum of the entropies of these macro states 
with the over all minus sign on as seen.
  \item In the case of time axis being compactified to $S^1$ the formula for the of 
``number of solutions" passing through the chain of macro states, 
which must now be a circular chain, is given again by selecting maximal and minimal entropy macro states as
\begin{eqnarray}
\exp\left(\sum _{\substack{i\in minimal ~S\\macro ~states}}S_{min.i}
-\sum _{\substack{j\in maximal ~S\\macro ~states}}S_{max.j}\right)
~. \label{3} 
\end{eqnarray}
This ``number of solutions" \cite{10} is meant to be a weighted number taking into account 
that the solutions are typically unstable in the sense that a tiny deviation makes 
it completely miss its periodicity as we shall discuss in the next section.

Since obviously this ``number of solutions" as given by (\ref{2}) or (\ref{3}) 
is at most unity, and that only when the entropy is constant along the solution path, 
we see that for the $S^1$-compactified time axis the entropy must be constant.
\end{enumerate}
\end{enumerate}

This is one of our main results in the present article that for compactified time $S^1$ 
and the above no fine tuning etc. assumptions entropy is forced to be constant as function of time.

This result means that since phenomenologically we know that after $\dot{S}>0$ 
we have excluded that the universe could indeed have an $S^1$ compactified time.
In case one should want to consider a model of that type one would have to defend it 
by pointing to errors in our assumptions or argumentation, or one would have to declare it 
in some way a fake that entropy increases. 

\section*{Effective number of solutions}

Let us first have in mind that imposing a given period for a given mechanical system -- 
in the general phase space formulation -- 
means restricting it by as many equation as there are dimensions in the phase space.
This is also the number of parameters needed to specify the initial state 
and thereby a solution to the Hamilton equations.
Thus by specifying the given period we have set up just equally many equations as 
are needed to fix the initial state and thus there will be no more parameters to specify a solution.
Rather there will be expected to be only a discrete set of solutions.

This is an important point of the present article to have in mind: 
Imposing the periodicity is essentially a replacement for assuming what the initial conditions are.
After having assumed the $S^1$-time axis there is no question any more of choosing initial conditions.
It is already done.

It should be admitted that when we have a system with Lyapunov exponent(s) non-zero, 
a typical chaotic system, the number of solutions with the prescribed period $T$ can easily 
become tremendous as we also estimated in.
Now, however, in such systems you also have that the solutions are typically ``unstable" 
in the sense that a solution which at one time $t_0$ is very close to a given solution will 
separate exponentially, i.e. proportional to 
$\exp(t\lambda_{Lyapunov})$, from the first solution.
Such an enhancement of any little deviation from the original solution means that 
if there for any reason -- say Heisenberg uncertainty principle -- 
is some uncertainty in the exact solution the probability for the slightly blurred point, 
i.e. then spot, in phase space will gradually leave further and further out.
Interestingly enough one easily sees that independent of the exact shape and extension 
of a spot around the classical solution considered, a spot of the percentage of such neighboring tracks 
remaining in the spot is calculable from the dynamics of the system.
Indeed it is obtained from the Taylor expansion up to second order of the Hamiltonian around the original solution.
Indeed from the matrix
\begin{eqnarray}
\underline{M}=\left(
  \begin{array}{cc}
\frac{\partial^2 H}{\partial q_{j} \partial p_i}       & 
\frac{\partial^2 H}{\partial p_{j} \partial p_i}   \\
\frac{\partial^2 H}{\partial q_{j} \partial q_i}       & 
\frac{\partial^2 H}{\partial p_{j} \partial q_i}    \\ 
\end{array}
\right)
\end{eqnarray}
one can extract Lyapunov exponents.
Here $N$ is the number of degrees of freedom and the four symbols 
$\frac{\partial^{2}H}{\partial q_{j}\cdot \partial q_{i}}$ etc. stand for the four $N\times N$ submatrices, 
with $i$ enumerating the rows and $j$ the columns.
>From $\underline{M}$ one may take the positive real eigenvalues --there are also imaginary ones $iw$ -- 
and consider them a sort of local ``Lyapunov exponents" 
giving how a cluster of solutions in the neighborhood of the original one 
are getting expanded in the  eigendirections.
This type of arguments lead easily to an average Lyapunov exponent defined by
\begin{eqnarray}
\gamma_{av}(t)=\sum _{\rm the~different~eigenvalues}|\gamma (t)|
~. \end{eqnarray}

Then the fraction of neighboring tracks lost per unit time is
\begin{eqnarray}
``Fraction ~lost/unit~time"
=\sum _{\gamma}|\gamma (t)|
= \gamma_{av}(t)
~. \end{eqnarray}

Integrating up this loss leads to the fact that the chance for a neighboring solution 
staying neighboring all through the prescribed period $T$ is
\begin{eqnarray}
&&\exp\left(-\int _{0}^{T} \gamma_{av}(t)dt\right)
=
%\nonumber\\&&
\exp\biggl(-\gamma_{av}(t)|_{\rm time ~averaged}  \cdot T \biggr)
~. \label{7}
 \end{eqnarray}
If for any reason there is a limited accuracy then a solution to the periodicity constraint 
should only be counted with the weight eq.(\ref{7}).
Now it was a point of the calculation of the Proceedings of the Bled conference \cite{6}
that this weighting actually canceled the number of solutions which were estimated 
to depend on the average Lyapunov exponents in just a compensating way.
This may not be so surprising since it is the same Lyapunov exponents that 
give the spread of the at first neighboring tracks over phase space thereby 
and which allows them to spread so as to give 
more and more solutions to the periodicity.

The end result is that taking into account both the weighting and that number 
of solutions to be found calculating with Lyapunov exponents gives no different statistical 
result compared to what we would get without them \cite{6}.

\section*{The likely entropy}
Considering the above formula for the number of solutions we see that this number 
becomes of order of unity for the entropy along the chain of entropies past in time being constant.
But this is our true main point: we get the same order of magnitude of the number of solutions 
from whatever the value of the constant entropy is!
All that matters in this formula is that it \underline{is} constant,
but the value it takes does not matter.

According to the jut mentioned arguments for the cancellation in the number of solutions 
counted with weight of the effect of the Lyapunov exponents we get of order one solution 
with such effects included whatever the constant entropy is, 
just provided that it is constant.

This is actually a remarkable result seen in the light of that if we considered a random micro state 
it would have the maximal entropy reachable.
But the above estimates say that the essentially unique solution with the intrinsic periodicity 
imposed does \underline{not} have typically this maximal entropy but rather could accidentally 
have whatever entropy, with an essentially smooth or flat probability distribution over entropy values is!

This result we conceive of as a half way success of compact time world in as far as one of the things 
to be explained is that the entropy is not maximal.
But of course since we also got the entropy constant as a function of time the predictions 
from such a closed circle time universe model are not in agreement with experiment.

Thus our compact time world model is not a viable model in spite of its partial success of at least 
not getting just the entropy maximal.

The partially successful latter result may be understandable by thinking of the system of 
a random walk inside some macro state or several such macro states.
In order to fulfill the given period $T$ restriction this random walk has to turn back 
after the period $T$ to its starting point.
Such a finding back will obviously have a probability proportional to the inverse phase space 
volume of the macro state in which it walks around, i.e. proportional to $e^{-S}$, where $S$ 
is the entropy of this macro state.
Now, however, you may imagine attempting to find a solution by starting at any (random) point 
in phase space.
That means that this randomly distributed test start has probability 
$\propto e^{S}$ for a macro state with entropy $S$.
Now as we saw the probability of successful return goes a $e^{-S}$ 
and therefore the probability for it being in a given macro state goes as 
$e^S \cdot e^{-S} \sim 1$ i.e. it is the same for them all.
 
\section*{Quantum Treatment}
In this article until now we have only used classical mechanics except for 
that we used the Heisenberg uncertainty relation as an argument for some uncertainty, 
but that could have been also some technical uncertainty.

An easy way to see our main result of entropy not likely maximal in a simple example 
is to consider that the total energy $E$ is one of the characteristics of the macro states:
Then in quantum mechanics it is assumed that all the quantum states in a given 
macro state corresponding to subspace have the energy $E$ as eigenvalue for the Hamiltonian operator $\hat{H}$.
That is to say if
\begin{eqnarray}
|a \rangle  \in ``Macro ~state ~A"
\end{eqnarray}
then
\begin{eqnarray}
\hat{H}|a \rangle= E_A |a\rangle
\end{eqnarray}
where $E_A$ is the energy characteristicum of the macro state $A$.

But the intrinsic periodicity with period $T$ means that
if the system passes through Macro state $A$ then
\begin{eqnarray}
E_{A}T = 2 \pi n, ~~~~~n=integer 
~.\label{10} 
\end{eqnarray}

This selection of macro states by such an integer relation (\ref{10}) involving the $a$ priori 
``random" in advance given period $T$ will normally not at all lead to a macro state with maximal entropy, 
but rather give a randomly macro states from which all sizes of entropy have an about equal probability.

This corresponds to the result we also obtained from classical consideration.

There is the possibility of developing a quantum analogue to the arguments we gave above and derive essentially 
the same rules quantum mechanically.

Analogously to the development of rules for how many states or how big phase space volume 
can come through a given series of macro state above we may formulate these rules quantum mechanical.
We can imagine that we have introduced projection operators for all the macro states 
$P_{A}, ~~A=1, 2, \ldots, n$, and 
in the chain of macro states discussed above and then the number of micro states coming 
through such a chain as defined above would be 
\begin{eqnarray}
&&\#states ~~through =
%\\&& 
Tr\left(\left(\prod _{A ~over ~the ~chain}P_{A}\right)
\left(\prod _{B ~over ~the ~chain}P_{B}\right)^{+}\right)
~.\nonumber 
\end{eqnarray}
Here we used -- following Gell-mann and Hartle -- the Heisenberg picture.
With Schr\H{o}dinger picture we would just stick in time-development operators
\begin{eqnarray}
\exp\left(-iH(t_{A}-t_{A-1})\right)
\end{eqnarray}
in between the factors so as to replace $\prod _{A}P_{A}$  by
\begin{eqnarray}
&&
e^{-i(t_{out}-t_{n})H}\cdot P_{n} \cdot 
\left(\prod _{A}^{n}(e^{-i(t_{A+1}-t_{A})H}P_{A})\right) \cdot 
%\nonumber\\&&~~~
e^{-iH(t_{1} - t_{start})}
\end{eqnarray}
Thus we have 
\begin{eqnarray}
&\#&``states ~~through" 
\nonumber\\
&=& 
%%\nonumber\\&&
Tr  \Bigg[
\bigg(e^{-i(t_{end}-t_{n})H}\cdot P_{n}e^{-iH(t_{n}-t_{n-1})}P_{n-1}\cdot 
%\nonumber\\&&~~~~~~~
e^{-iH(t_{n-1}-t_{n-2})} \cdot P_{n-2} \cdots  \cdot 
P_{1} \cdot e^{-iH(t_{1}-t_{start})} \bigg) \cdot 
\nonumber\\&&~~~~~~~
\bigg( e^{-i(t_{end}-t_{n})H} \cdot P_{n} \cdot \cdots \cdot 
P_{2} \cdot
%\nonumber\\&&
 e^{-iH(t_{2}-t_{1})} \cdot P_{1} \cdot 
e^{-iH(t_{1}-t_{star})} \bigg)^{\dag} \Bigg]
\end{eqnarray}
in the Schr\H{o}dinger formulation.
Here $t_A$ is a time $t$ representing the era of the macro state $A$ 
(with entropy $S_A$) and $t_{end}$ and $t_{start}$ are the formal end and start times.

Of course in this formalism analogously to the phase space volume we have entropy definitions above 
in the classical case, now simply
\begin{eqnarray}
S_{A} = \log Tr(P_{A})
\end{eqnarray}
where we used the usual also normalizing equations for projection operators
\begin{eqnarray}
P_{A}^{2} = P_{A} = P_{A}^{+}
~.\end{eqnarray}

Corresponding to the assumption 
A) above of neighboring macro states in the chain being ``maximally coupled" 
we would have in say the Schr\H{o}dinger picture notation
\begin{eqnarray}
P_{A} \cdot e^{-i(t_{A}-t_{B})H} \cdot P_{B}=
e^{-i(t_{A}-t_{B})H} \cdot P_{B}
\end{eqnarray}
when $S_{B}<S_{A}$.

We should also have in mind that when the system is supposed to be in some macro state $C$ 
then the Hamiltonian is for our purposes a random one only 
though not sending the system out of that macro state $C$.
This means that in the era of $C$ we effectively have
\begin{eqnarray}
[H, ~P_{C}] = 0
\end{eqnarray}
but will have random terms in $H$ making
\begin{eqnarray}
[H, ~P_{A}] \neq  0
.~\end{eqnarray}

One can then easily derive expressions corresponding to our above rules.

For the case of compactified time we have
\begin{eqnarray}
&&
\#``states ~~through" =
%\nonumber\\&&
\left| Tr\left(\prod _{A}(e^{-i(t_{A+1}-t_{A})H}P_{A})\right)
\right| ^{2}
~.
\end{eqnarray}
As is easily seen one gets the expressions from above.

\section*{Conclusion and outlook}
In the present article we have put forward some estimation procedures for the number of solutions 
of the classical equations of motion or the thereby proportional phase space volume of 
such solutions by means of the entropies of those macro states through which the to be 
counted solutions are required to pass at different times.
Most importantly we like in the Proceedings of the Bled Conference \cite{6} that for a 
compact $S^{1}$ time axis the solution to the equations of motion gets essentially fixed, 
and that extremely likely not to the maximal entropy.
Rather the entropy can get whatever value is.
However, it follows also that the entropy becomes constant as a function of time.
When we here used the word ``essentially" about the fixed unique solution, it first refers to 
that the solution need not be totally unique although a discrete set up to the trivial time 
translation one parameter set of solutions; 
but even apart from the time translation it is only the ``weighted" number that is really of 
order unity even.
If the solutions have non-zero Lyapunov exponents there will be very many solutions, 
but they will be ``unstable" in the sense that a tiny little error would grow up exponentially 
and thus if there is some uncertainty in the track -- like Heisenberg uncertainty say -- 
the solution will effectively die away exponentially with time along the path.
Including this effect by taking the weighting estimating the chance that a given solution 
even with error return as periodic, we get a compensation for the increased number of mathematical solutions.
The weighted number of solutions become again of order unity.

As further outlook let us mention that we have already attempted to avoid the problem of constant entropy 
in the $S^{1}$-time model
by considering in \cite{11}
a model with instead an infinite time axis $(-\infty, \infty )$.

Without the compactification we have, however, no immediate fixing of the actual solution to the equation of motion 
-- the initial conditions so to speak -- 
and we had to help on that problem to introduce a probability density $P(path)$ 
over the set of possible solutions of the equation of motion.
This game turned out much more successfully phenomenologically and actually constitutes 
a viable model, even and this is surprising, with the general principles of the other laws, 
other than second law of thermodynamics such as locality in time, various symmetries, 
even including time-reversal invariance.

The latter seems of course a priori to be almost doomed to be catastrophic 
if you hope to get even only an effective second law of thermodynamics out.
%%%%% END %%%%%

%\section*{Conclusion}

\begin{acknowledgments}
This work is supported by Grant-in-Aid for Scientific Research on Priority Areas, 
Number of Area 763 ``Dynamics of Strings and Fields" 
from the Ministry of Education of Culture, Sports, Science and Technology, Japan.
\end{acknowledgments}

\end{document}